     \documentstyle[12pt,epsfig,amsfonts]{article}
     \newlength{\dinwidth}                       
     \newlength{\dinmargin}                      
\def\lsim{\mathrel{\rlap{\lower4pt\hbox{\hskip1pt$\sim$}}
    \raise1pt\hbox{$<$}}}                
\def\gsim{\mathrel{\rlap{\lower4pt\hbox{\hskip1pt$\sim$}}
    \raise1pt\hbox{$>$}}}                
\def\be{\begin{equation}}
\def\ee{\end{equation}}
\def\bea{\begin{eqnarray}}
\def\eea{\end{eqnarray}}

\def\bit{\begin{itemize}}
\def\eit{\end{itemize}}

\def\h5{\hskip5mm}

\newcommand{\alps}{\mbox{$\alpha_s$ }}

\newcommand{\GeV}{\,\mbox{GeV}}

\def\cm{\checkmark}

\newcommand{\err}{\,$\pm$\,}
\begin{document}
\hfill{\parbox[t]{4cm}{MADPH-99-1129  \\ PITHA 99/30 \\ hep-ph/9910448 }}
\vspace*{5mm}
%
%
\begin{center}  \begin{Large} \begin{bf}
Comparison of Next-to-Leading Order Calculations \\
for Jet Cross Sections in Deep-Inelastic Scattering\\
  \end{bf}  \end{Large}
  \vspace*{5mm}
  \begin{large}
 C. Duprel$^a$, Th. Hadig$^b$, N. Kauer$^c$, M. Wobisch$^a$ \vskip2mm
  \end{large}
\end{center}
$^a$ III. Physikalisches Institut, RWTH Aachen, D-52056 Aachen, Germany \\
$^b$ I. Physikalisches Institut, RWTH Aachen, D-52056 Aachen, Germany \\
$^c$ Department of Physics, University of Wisconsin, Madison, WI 53706, USA \\
\begin{quotation}
\noindent
{\bf Abstract:}
We compare different next-to-leading order calculations
of jet cross sections in deep-inelastic scattering as implemented
in the programs DISASTER++, DISENT, JETVIP and MEPJET.
In all phase space regions under study DISENT and DISASTER++ agree 
better than 2\%.
MEPJET shows systematic deviations of being typically 5--8\% lower
than the other programs.
The JETVIP results show a significant dependence on the
phase space slicing parameter $y_{\rm cut}$.
In the cases where the $y_{\rm cut}$ dependence within 
$10^{-4} \le y_{\rm cut} \le 10^{-3}$ is smaller than 3\% the JETVIP 
results are often comparable with the DISENT and DISASTER++ results.
\end{quotation}

\section{Introduction}
At present four different Monte Carlo programs are available for 
the computation of 
jet quantities in deep-inelastic scattering (DIS) to
next-to-leading order (NLO) in the strong coupling constant $\alpha_s$:
MEPJET~\cite{mepjet,mirkeshabi}, DISENT~\cite{disent}, 
DISASTER++~\cite{disaster} and JETVIP~\cite{jetvip}.
Since all of these claim to be exact calculations, they should produce 
identical results (within numerical precision).
In this contribution we compare the leading-order (LO) and the NLO 
predictions of these programs to test whether
they are compatible.
The comparisons are performed in typical phase space regions 
where HERA analyses are currently made.

\section{Program Overview}
The four programs allow to calculate next-to-leading order parton 
cross sections with arbitrary cuts.
They differ in the techniques used and in several details.
A short overview on the four programs is given in
table~\ref{tabfeature}. 
For a detailed discussion of the single topics
we refer to the program manuals.

\begin{table}[htb]
{\small
\begin{tabular}{@{}lllll}
\hline
&MEPJET\cite{mepjet} &DISENT\cite{disent} &DISASTER++\cite{disaster} 
&JETVIP\cite{jetvip} \\
\hline
version & 2.2 & 0.1 & 1.0.1 & 1.1\\
\hline
method & PS slicing & subtraction & subtraction & PS slicing \\
\hline
1+1,2+1 & NLO     & NLO    & NLO    & NLO \\
3+1     & LO      & LO     & LO     & LO \\
4+1     & LO      & ---    & ---    & --- \\
\hline
full event record & \cm & \cm & \cm & (\cm) \\
\hline
scales       & all    & factorization: $Q^2,$ fixed & all & all \\
             &        & renormalization: all &  & \\
\hline
flavour dependence & switch & switch & full & switch \\
\hline
quark masses &    &    &   & \\ 
\, \, \, in LO x-section    & LO     & ---    & ---  & --- \\
resolved $\gamma$ contribution  & &  & &  \\ 
\, \, \, in LO/NLO x-section            & --- & --- & --- & NLO \\
electroweak contribution  &    &    &   & \\ 
\, \, \, in LO/NLO x-section    & LO & --- & --- & --- \\
polarized x-section  & NLO & --- & --- & --- \\
\hline
\end{tabular}
\caption[Comparison of the different features of the programs.]
{{\it Comparison of the different features of the programs. Note that DISENT
has been changed with respect to the official version to implement 
e.g.\ the running electromagnetic coupling constant. The 
common NLO library~\cite{commonlib} version 0.2 has been used
to interface DISASTER++.}}
\label{tabfeature}
}
\end{table}

The programs can be classified by the method that is
applied to cancel the collinear and infrared singularities.
Two general methods are available, the phase space slicing
method and the subtraction method.  
The phase space slicing
method employs a technical cutoff ($s_{\rm min}$ or $y_{\rm cut}$).  
Correct results are only obtained for sufficiently small values of
this parameter.  
The cutoff independence has to be checked for
every investigated observable/scenario.  
In practice this test is performed by comparing multiple runs 
with different (small) cutoff parameters.  
The subtraction method does not apply such a cutoff.

All programs are able to calculate single jet and dijet observables in 
next-to-leading order, i.e.\ ${\cal O}(\alpha_s^1)$ 
or ${\cal O}(\alpha_s^2)$ for processes with one or two partons in the
Born process.
Processes with a higher number of particles in the Born graph are
available in leading order only.

In order to apply arbitrary cuts on the final state, the full event record of
all incoming and outgoing particles is needed. 
The full event record is available for all 
programs with the exception of the azimuthal angle $\phi$ of the scattered
electron wrt.\   the outgoing partons in the JETVIP program.
In JETVIP the $\phi$ dependence of the matrix elements is integrated
analytically.
Since this angle is not available, the full vector of the 
Lorentz boost from the Breit frame to the HERA laboratory frame 
can only be calculated under the assumption of a flat distribution
in $\phi$.
At larger $Q^2$ this can lead to an error of at maximum
5-7\% when angular jet cuts in the HERA laboratory frame are 
applied~\cite{mirkeshabi}.
Therefore no such cuts are used in our test scenarios.

In perturbative QCD calculations two scales are introduced:
the renormalization ($\mu_r$) and the factorization scale ($\mu_f$). 
All programs allow to identify the renormalization scale with 
arbitrary varaibles,
e.g.\  proportional to kinematic variables ($Q$)
or to final state quantities ($E_{T{\rm ,\, jet}}$).
The same is true for the factorization scale, except for DISENT.
In DISENT the factorization scale is restricted to variables 
that are independent of the hadronic final state,
i.e.\  proportional to kinematic variables ($Q$) or to constant values.
To keep the checks simple, we stick to the choice of $\mu = Q$ 
for both scales.

\newpage

At very low and at very high $Q^2$, effects changing the cross section
become more and more important. 
At high $Q^2$ the exchange of $Z$ and $W$ bosons can not be neglected 
while at low $Q^2$ the contributions from resolved photons
to jet cross sections become sizable.
In other regions of phase space effects from quark masses can 
also become relevant.
Since these different effects can only be calculated by single 
programs (see table) they have not been considered in the present
comparison.

\section{Comparison of the Results \label{secscenarios}}

\subsection{Technical Settings}
For all NLO calculations as well as for the LO calculations we are using
renormalization and factorization scales $\mu_r = \mu_f = Q$
and the 2-loop formula for the running of \alps 
(taken from PDFLIB~\cite{pdflib}).
Throughout we are using CTEQ4M parton density functions~\cite{cteq4m} 
(taken from PDFLIB).
All cross sections are calculated for a running electromagnetic coupling
constant\footnote{The official DISENT program does not take into
account the running of the electromagnetic coupling constant.
We have modified the official DISENT program to include this.}
and are performed in the $\overline{\rm MS}$ scheme with five 
active flavors.

\subsection{Scenarios for the Comparisons}
At NLO the jet cross sections depend on the jet definition 
and the recombination scheme.
For all comparisons we are using the inclusive $k_\perp$
algorithm~\cite{inclkt} in the Breit frame.
It has been shown that this jet definition is infrared safe to all
orders~\cite{seymour98} and less affected by hadronization
corrections than other jet definitions~\cite{wobischwengler}.
Particles are recombined in the $E_T$-scheme~\cite{etscheme}
in which the jet $E_T$ is obtained from the scalar sum of the particle 
$E_T$, the pseudorapidity and the azimuth angle are calculated as
$E_T$-weighted averages from the particle quantities.
In all cases we calculate {\em inclusive} jet cross sections
(i.e. cross sections for the production of events with at least two
jets that pass the jet cuts).
The jets are indexed in descending order in their transverse energies
in the Breit frame ($E_{T1} \ge E_{T2}$).

The $ep$ center of mass energy squared is set to 
$s = 4 \cdot 27.5 \cdot 820\GeV^2 = 90200\GeV^2$
(corresponding to the HERA running conditions in 1994-97).
What will later be called the ``central scenario'' is defined as
follows:
\be
30 < Q^2 < 40\GeV^2 \, ,  \hskip9mm    0.2 <  y < 0.6 \, ,
\hskip9mm E_{T2{\rm min}} = 5\GeV \, .
\ee
So far this scenario includes infrared sensitive parts of phase space, 
where $E_{T1} \simeq E_{T2} \simeq E_{T{\rm min}} $.
These phase space regions can be avoided by additional harder cuts 
on either the $E_{T1}$ of the hardest jet, on the average 
$\overline{E}_T$ of both jets or on the invariant dijet mass $M_{jj}$.
These different choices are varied in scenario 1(a-c).
The central choice will be an additional cut on $E_{T1} > 8\GeV$.
Different values of this cut are tested in scenario 2(a-d).

Starting from the central scenario we also vary the ranges
of the kinematical variables $Q^2$ (scenario 3(a-d)) and 
$y$ (scenario 4(a-c)).
Further comparisons are dedicated to phase space regions 
which are irrelevant for experimental analyses, but helpful 
to test the programs.
In scenario 5(a-c) we compare the programs for softer transverse 
jet energy cuts.
Scenario 6(a-c) is defined by the requirement of a difference
in the transverse jet energies. 
The only contributions to these cross sections come from 
3-parton final states in ${\cal O}(\alpha^2_s)$ such that
we are left with a leading order prediction.

The various scenarios differ from the central scenario (1)
as follows: \newline

\hskip10mm
\begin{tabular}[b]{r|c}
\multicolumn{2}{c}{\sc Scenario 1}   \\
\hline
\hline
\multicolumn{2}{l}{different ways to avoid} \\
\multicolumn{2}{l}{infrared sensitive regions} \\
\hline
No. & additional jet cut \\
\hline
1 a) &  $E_{T1{\rm min}} > 8\GeV$ \\
1 b) &  $M_{jj} > 25\GeV$  \\
1 c) &  $(E_{T1} + E_{T2}) > 17\GeV$     \\
\hline
\end{tabular}
\hskip29mm
\begin{tabular}[b]{r|r}
\multicolumn{2}{c}{\sc Scenario 2}   \\
\hline
\hline
\multicolumn{2}{l}{different  $E_{T1{\rm min}}$ cuts } \\
\hline
No. & $E_{T1{\rm min}} / \GeV$\\
\hline
2 a) &    8 \\
2 b) &   15 \\
2 c) &  25 \\
2 d) &  40 \\
\hline
\end{tabular}

\vskip10mm
\hskip10mm
\begin{tabular}[b]{r|r|r}
\multicolumn{3}{c}{\sc Scenario 3}   \\
\hline     \hline
\multicolumn{3}{l}{different $Q^2$ ranges}   \\
\multicolumn{3}{l}{add.\   cut $E_{T1{\rm min}} = 8\GeV$}   \\
\hline
No. & $Q^2_{\rm min} / \GeV^2$ &  $Q^2_{\rm max} / \GeV^2$ \\
\hline
3 a) &    3  &   4 \\
3 b) &   30  &   40 \\
3 c) &  300  &  400 \\
3 d) & 3000  & 4000  \\
\hline
\end{tabular}
\hskip23mm
\begin{tabular}[b]{r|l|l}
\multicolumn{3}{c}{\sc Scenario 4}   \\
\hline
\hline
\multicolumn{3}{l}{extreme $y$ regions} \\
\multicolumn{3}{l}{add. cut $E_{T1{\rm min}} = 8\GeV$} \\
\hline
No. & $y_{\rm min}$ & $y_{\rm max}$ \\
\hline
4 a) &    0.01 & 0.05 \\
4 b) &   0.2 & 0.6 \\
4 c) &  0.9 & 0.95 \\
\hline
\end{tabular}

\vskip10mm
\hskip10mm
\begin{tabular}[b]{r|r|r}
\multicolumn{3}{c}{\sc Scenario 5}   \\
\hline
\hline
\multicolumn{3}{l}{probing softer regions} \\
\hline
No. & $E_{T2{\rm min}} / \GeV$  & $E_{T1{\rm min}} / \GeV$ \\
\hline
5 a) & 1 & 2 \\
5 b) & 2 & 3 \\
5 c) & 3 & 4  \\
\hline
\end{tabular}
\hskip20mm
\begin{tabular}[b]{r|c}
\multicolumn{2}{c}{\sc Scenario 6}   \\
\hline
\hline
\multicolumn{2}{l}{add. cut on the difference} \\ 
\multicolumn{2}{l}{of the jet $E_T$} \\
\hline
No. &  $(E_{T1} - E_{T2}) >$ \\
\hline
6 a) &  1 GeV \\
6 b) &  2 GeV \\
6 c) &  3 GeV \\
\hline
\end{tabular}

\vskip15mm

\subsection{Numerical Comparisons} 
An overview of the results of all calculations for the 17 different
scenarios is given in the tables in the appendix.
The leading order results are shown in the last row for each scenario.
These values have been calculated to a numerical precision of 
typically 0.2\%. 
In all cases we see a perfect agreement between the different programs.

The next-to-leading order calculations for the corresponding scenarios
have been performed to a numerical precision of typically
0.3\% \footnote{For DISASTER++ the precision is often worse
since the calculations by DISASTER++ require 
significantly more CPU time compared to the other programs.}.
In most cases we have tested the stability of the JETVIP results 
w.r.t.\ the cutoff parameter $y_{\rm cut}$.

\subsubsection*{DISENT and DISASTER++}
The programs DISENT and DISASTER++ which are both based on the
subtraction method are in very good overall agreement.
In 12 cases their NLO results are in agreement within the quoted 
errors of typically 0.3\%.
Only in 5 comparisons (1b, 3c, 3d, 4a, 5a) deviations are seen
in the range of 0.6\% to 2.2\% with a significance of 1.5 -- 3.6 standard 
deviations but without any systematic trend.
If we consider that the errors quoted by the programs may 
sometimes be 
underestimated\footnote{For the DISENT program the same cross section
has been repeatedly calculated~\cite{seymourpriv}. 
The statistically independent results were roughly Gaussian distributed 
in the central region, with a width compatible with the error quoted by 
DISENT.
However, significantly larger tails have been seen.
The same is likely to be true for the other programs
(no similar checks have been made).}
this can still be labeled ``good agreement''.

In cases where the precision of the NLO calculation is important, 
the user should therefore not trust the quoted erros but 
aim for a higher precision.

\subsubsection*{MEPJET} 
MEPJET NLO predictions were found to agree well with 
DISASTER++ and DISENT for physical PDFs in Ref.~\cite{disaster} 
(Table 2 and discussion on p.~14).  
MEPJET's NLO results for the extreme phase space regions 
investigated here are typically 5--8\% lower than the NLO results 
obtained with DISENT and DISASTER++.  
In three cases deviations of about 10\% occur.  
No obvious correlation between the size of the deviation and the 
value of the $K$-factor exists.  
The $K$-factor varies from 1.3 to 7.0, except for case 2d.  
Here DISASTER++ and DISENT yield a $K$-factor of 1.17 and 1.16, 
respectively, while MEPJET's $K$-factor is close to 1.
MEPJET deviates from DISASTER++ by 2$\sigma$ and from DISENT by
3$\sigma$ in this case.

All MEPJET calculations ran with the default cutoff value of 
$s_{\rm min} = 0.1\GeV^2$.
To check for possible cutoff dependences
additional runs with smaller $s_{\rm min}$ values were done for selected
cases (see data table).  No $s_{\rm min}$ dependence was found.  
Effects potentially introduced by approximations used for the 
crossing functions were also investigated and found to be not significant.  
All LO results agree within the statistical errors.  
In addition perfect agreement between MEPJET, DISASTER++
and DISENT is seen in scenario 6, which tests the real $O(\alpha_s^2)$
corrections.
What causes the observed discrepancies in full NLO in the extreme phase
space regions probed here is currently unknown.

\subsubsection*{JETVIP}
As proposed in~\cite{jetvip} we have started to perform the NLO 
calculations for the JETVIP program for a cutoff value 
$y_{\rm cut} = 10^{-3}$.
Although some of these results are in agreement to
the DISENT/DISASTER++ values (scenarios 2b, 3d, 4a, 5a-c),
in the other 11 cases discrepancies of up to 20\%
are seen.
Therefore we have made extensive studies on the $y_{\rm cut}$
dependence of the JETVIP results in the range 
$10^{-6} \le y_{\rm cut} \le 10^{-2}$.
Only in scenario 6, where only real corrections of
${\cal O}(\alpha_s^2)$ are tested, the results become stable 
for $y_{\rm cut} \simeq 10^{-4}$.
For all NLO results we observe a significant cutoff dependence.
Since the independence on the cutoff is the most important
test of the successful implementation of the phase space slicing
method the strong $y_{\rm cut}$ dependence of the JETVIP results
is worrisome.

Especially at very small values of $10^{-6} \le y_{\rm cut} \le 10^{-5}$
no convergence of the results is seen.
In scenario 1a we have repeated the calculation at
$y_{\rm cut} = 10^{-5}$ with fourfold statistics.
While the quoted errors are 2.6\% and 0.4\%, respectively,
both results deviate by 15\%.
This is a clear indication that at these small $y_{\rm cut}$ values 
the quoted errors are not reliable.

At intermediate values $10^{-4} \le y_{\rm cut} \le 10^{-3}$
large $y_{\rm cut}$ dependencies (above 4\%) are observed  
in four scenarios (2c, 2d, 3a, 6a) only.
In the other 13 scenarios the dependence is below 3\%.
In 11 of these cases the JETVIP results at $y_{\rm cut} = 10^{-4}$
agree within this level of precision with the DISENT/DISASTER++ 
results. 
The other two results 1b, 1c deviate by 10\% and 4.5\% from
the DISENT/DISASTER++ results.

\section{Summary}

We have compared
four different programs for NLO calculations
of jet cross sections in $ep$ collisions:
DISENT, DISASTER++, JETVIP and MEPJET.
Dijet cross sections in different ranges of $Q^2$, $y$, $E_{T{\rm ,\, jet}}$
have been calculated in leading order (LO) and in next-to-leading order (NLO).
All calculations are performed to a numerical precision of typically
0.2\% (LO) and 0.3\% (NLO).

While the leading order predictions of all programs agree
within the numerical precision of 0.2\%,
our comparisons show that in NLO only the calculations of 
DISENT and DISASTER++ can be said to be in good agreement.

MEPJET shows systematic deviations of being typically 5--8\% lower
than DISENT and DISASTER++.
Only the ${\cal O}(\alpha_s^2)$ tree level cross sections
are in perfect agreement.

The JETVIP program shows a significant dependence on the
phase space slicing parameter $y_{\rm cut}$ which has to be 
understood.
Only at intermediate values of $y_{\rm cut} \simeq 10^{-4}$
the dependence is reduced.
In the cases where the $y_{\rm cut}$ dependence  within 
$10^{-4} \le y_{\rm cut} \le 10^{-3}$ is smaller than 3\% the JETVIP 
results are often comparable with the DISENT and DISASTER++ results.

\vskip15mm

\paragraph{Acknowledgments}
We would like to thank Stefano Catani, Dirk Graudenz, Erwin Mirkes, 
Bj\"orn P\"otter, Mike Seymour, Dieter Zeppenfeld for providing the
NLO programs and for many helpful discussions.

\newpage


\newpage

\begin{appendix}
\section{Numerical Results}
Here we list all available numerical results.
The last line for each scenario contains the leading-order results.

\vskip7mm

\noindent
{\footnotesize
\begin{tabular}{c|r|r|r|r}
\hline
scenario  &  DISASTER++ & DISENT & JETVIP & MEPJET \\
\hline
\hline
1 a) & 119.82 \err 0.411 &  119.54 \err 0.33 &   113.42 \err 0.10 ($y_{\rm cut}=10^{-2}$) & 113.45 \err 0.21 ($s_{\rm min}=0.1$) \\
  & & & 121.41 \err 0.19 ($y_{\rm cut}=10^{-3}$) &
 113.3 \err 3.5 ($s_{\rm min}=0.0001$) \\
  &  & & 121.69 \err 0.77 ($y_{\rm cut}=10^{-4}$) &   \\
  &  & & 99.6 \err 2.6 ($y_{\rm cut}=10^{-5}$) &   \\
  &  & & 114.98 \err 0.44  ($y_{\rm cut}=10^{-5}$) &   \\
  &  & & 75.7 \err 2.7 ($y_{\rm cut}=10^{-6}$) &   \\
LO:      & 41.662 \err 0.083 &  41.769 \err 0.061 &  41.745 \err 0.033 & 
41.722 \err 0.032 \\
\hline
1 b) &  82.83 \err 0.44 & 81.02 \err 0.49 & 
93.58 \err 0.22 ($y_{\rm cut}=10^{-3}$) &  78.55 \err 0.16 \\
 & & & 91.11 \err 0.49 ($y_{\rm cut}=10^{-4}$) & \\ 
 & & & 83.55 \err 1.1 ($y_{\rm cut}=10^{-5}$) & \\ 
LO: & 30.57 \err 0.07 & 30.59 \err 0.05 & 30.54 \err 0.05 & 30.56 \err 0.01 \\
\hline
1 c) & 72.26 \err 0.30 & 72.05 \err 0.28 & 
77.15 \err 0.16 ($y_{\rm cut}=10^{-3}$) &  67.57 \err 0.21 \\
  & & & 75.46 \err 0.37 ($y_{\rm cut}=10^{-4}$) &     \\
  & & & 66.18 \err 1.36 ($y_{\rm cut}=10^{-5}$) &     \\
LO:    & 35.155 \err 0.072 & 35.172 \err 0.052 & 35.184 \err 0.050 &
 35.141 \err  0.024    \\
\hline
\hline 
\end{tabular}
}

\vskip9mm

\noindent
{\footnotesize
\begin{tabular}{c|r|r|r|r}
\hline
scenario  &  DISASTER++ & DISENT & JETVIP  & MEPJET \\
\hline
\hline
2 a) & as 1 a)    &     &    &     \\ 
\hline
2 b) & 16.585 \err 0.092 & 16.526 \err 0.051 & 16.668 \err 0.031 ($y_{\rm cut}=10^{-3}$) 
&  15.743 \err 0.078  \\
  &  &  & 16.302 \err 0.071 ($y_{\rm cut}=10^{-4}$) &  \\
  &  &  & 13.668 \err 0.160 ($y_{\rm cut}=10^{-5}$) &  \\
LO:      & 6.185 \err 0.020 & 6.222 \err 0.011 & 6.214 \err 0.005   & 6.221 \err 0.003    \\
\hline
2 c) & 2.0809 \err 0.0273 & 2.0519 \err 0.0080 & 1.9563 \err 0.0049 ($y_{\rm cut}=10^{-3}$) 
& 1.9084 \err 0.0083 \\
   & & & 1.7987 \err 0.0119 ($y_{\rm cut}=10^{-4}$) &  \\
   & & & 1.0962 \err 0.0260 ($y_{\rm cut}=10^{-5}$) &  \\
LO:      & 1.0230 \err 0.0046 & 1.0221 \err 0.0022 & 1.0255 \err 0.0009 & 1.0250 \err 0.0005 \\
\hline
2 d) & 0.1398 \err 0.0052 & 0.1403 \err 0.0011 & 0.1124 \err 0.0007 ($y_{\rm cut}=10^{-3}$) 
& 0.1229 \err 0.0047 \\
  &  & & 0.0772 \err 0.0014  ($y_{\rm cut}=10^{-4}$) &  \\
 LO:     & 0.1197 \err 0.0014 & 0.12125 \err 0.00036 & 0.12073 \err 0.00016 &
 0.12087 \err   0.000064    \\
\hline
\hline 
\end{tabular}
}

\vskip9mm

\noindent
{\footnotesize
\begin{tabular}{c|r|r|r|r}
\hline
scenario  &  DISASTER++ & DISENT & JETVIP & MEPJET \\
\hline
\hline
3 a) & 341.2 \err 1.7 & 339.1 \err 1.2
    & 315.9 \err 0.4 ($y_{\rm cut}=10^{-3}$)  &  331.49 \err 0.42 ($s_{\rm min}=0.1$) \\
        & & & 340.0 \err 0.7 ($y_{\rm cut}=10^{-4}$) & 334.96 \err 1.31 ($s_{\rm min}=0.01$) \\
    & & & 296.6 \err 2.0 ($y_{\rm cut}=10^{-5}$)  &  336 \err 14 ($s_{\rm min}=0.001$) \\
LO:      & 48.418 \err 0.100 & 48.423 \err 0.081   & 48.363 \err 0.040 &
 48.397 \err 0.040 \\ 
\hline
3 b) & as 1 a) & & & \\
\hline
3 c) & 26.848 \err 0.061 &  26.680 \err 0.051 & 
26.259 \err 0.139  ($y_{\rm cut}=10^{-3}$) & 24.684 \err 0.050 \\
  &  & & 26.79 \err 0.094 ($y_{\rm cut}=10^{-4}$) &   \\
  &  & & 23.894 \err 0.407 ($y_{\rm cut}=10^{-5}$) &   \\
LO:      & 16.938 \err 0.022 &  16.936 \err 0.016 & 16.928 \err 0.008  &
 16.918 \err 0.011    \\
\hline
3 d) & 1.9975 \err 0.0033 & 1.9852 \err 0.0029 & 
1.9657 \err 0.0061 ($y_{\rm cut}=10^{-3}$) & 1.8917 \err 0.0038 \\
  &  & & 1.9946 \err 0.0066 ($y_{\rm cut}=10^{-4}$) &   \\
  &  & & 1.7194\err 0.0179 ($y_{\rm cut}=10^{-5}$) &   \\
LO:      & 1.4982 \err 0.0017 & 1.4967 \err 0.0013 & 1.4956 \err 0.0013 & 
1.4966 \err 0.0010 \\
\hline
\hline 
\end{tabular}
}

\vskip9mm

\noindent
{\footnotesize
\begin{tabular}{c|r|r|r|r}
\hline
scenario  &  DISASTER++ & DISENT & JETVIP & MEPJET \\
\hline
\hline
4 a) & 19.218 \err 0.143 & 18.959 \err 0.068 & 18.818 \err 0.051 ($y_{\rm cut}=10^{-3}$) & 
17.190 \err 0.037  \\
 & & & 18.470 \err 0.041 ($y_{\rm cut}=10^{-4}$) & \\ 
 & & & 18.896 \err 0.167 ($y_{\rm cut}=10^{-5}$) & \\ 
LO:      & 11.611 \err 0.038 & 11.573 \err 0.022 & 11.590 \err 0.010 &
 11.587 \err 0.006    \\ 
\hline
4 b) &  as 1 a)   &    &    &     \\
\hline
4 c) & 6.424 \err 0.027 & 6.448 \err 0.018 & 6.299 \err 0.059 ($y_{\rm cut}=10^{-3}$) &
 6.086 \err 0.028  \\
 & & & 6.356 \err 0.040 ($y_{\rm cut}=10^{-4}$) & \\ 
 & & & 6.243 \err 0.155 ($y_{\rm cut}=10^{-5}$) & \\ 
LO:      & 2.1612 \err 0.0058 & 2.1615 \err 0.0031 & 2.173 \err 0.013 &
 2.1598 \err  0.0015    \\
\hline
\hline 
\end{tabular}
}

\vskip9mm

\noindent
{\footnotesize
\begin{tabular}{c|r|r|r|r}
\hline
scenario  &  DISASTER++ & DISENT & JETVIP & MEPJET \\
\hline
\hline
5 a) & 1676.2 \err 4.0 & 1655.6 \err 4.2 & 1654.3 \err 6.0 ($y_{\rm cut}=10^{-3}$) & 1489.8 \err 3.2    \\ 
 & & & 1678.6 \err 20.4  ($y_{\rm cut}=10^{-4}$) & \\ 
LO:      & 845.40 \err 1.04 & 844.71 \err 0.70 & 844.84 \err 0.83 & 
844.67 \err 0.45    \\ 
\hline
5 b) & 973.8 \err 2.6 & 970.1 \err 2.4 & 970.3 \err 3.0 ($y_{\rm cut}=10^{-3}$) &  885.9 \err 2.0  \\
 & & &  989.4 \err 8.3 ($y_{\rm cut}=10^{-4}$) & \\ 
LO:      & 436.43 \err 0.62 & 436.25 \err 0.43 & 436.85 \err 0.68 & 
436.27 \err 0.23    \\
\hline
5 c) & 564.5  \err 1.6  & 561.9 \err 1.5  & 
565.6 \err 1.5 ($y_{\rm cut}=10^{-3}$) & 518.0 \err 0.8 \\
 & & & 573.6 \err 4.8 ($y_{\rm cut}=10^{-4}$) & \\ 
LO:      & 242.20 \err 0.37 & 242.60 \err 0.28 & 243.25 \err 0.36 &
 242.47 \err 0.23    \\
\hline
\hline 
\end{tabular}
}

\vskip9mm

\noindent
{\footnotesize
\begin{tabular}{c|r|r|r|r}
\hline
scenario  &  DISASTER++ & DISENT & JETVIP & MEPJET \\
\hline
\hline
6 a) & 126.24 \err 0.44 & 126.92 \err 0.47 & 118.13 \err 0.05 
($y_{\rm cut}=10^{-3}$)  & 126.08 \err    0.20\\ 
      &    &  & 122.94 \err 0.05 ($y_{\rm cut}=10^{-4}$) & \\
      &    &  & 123.01 \err 0.05 ($y_{\rm cut}=10^{-5}$) & \\
      &    &  & 123.23 \err 0.22  ($y_{\rm cut}=10^{-6}$) & \\
\hline
6 b) & 56.30 \err 0.26 & 56.02 \err 0.25 &
 54.78 \err 0.02 ($y_{\rm cut}=10^{-3}$)  & 55.90 \err  0.10 \\
  &    &  & 55.30 \err 0.03 ($y_{\rm cut}=10^{-4}$)  &  \\
  &    &  & 55.30 \err 0.03 ($y_{\rm cut}=10^{-5}$)  &  \\
  &    &  & 55.30 \err 0.03 ($y_{\rm cut}=10^{-6}$)  &  \\
\hline
6 c) & 27.15 \err 0.16 & 27.13 \err 0.07 & 
26.94 \err 0.01 ($y_{\rm cut}=10^{-3}$) & 27.14  \err    0.05 \\
  &    &  & 27.00 \err 0.02 ($y_{\rm cut}=10^{-4}$)  &  \\
  &    &  & 27.00 \err 0.02 ($y_{\rm cut}=10^{-5}$)  &  \\
  &    &  & 27.01 \err 0.02  ($y_{\rm cut}=10^{-6}$)  &  \\
\hline
\hline 
\end{tabular}
}

\end{appendix}

\end{document}